\documentclass{bioinfo}
\copyrightyear{2015} \pubyear{2015}
\access{Advance Access Publication Date: 20 Nov 2021}
\appnotes{Article}
\usepackage{mathtools}
\usepackage{booktabs}
\usepackage{amsfonts}
\usepackage[para,online,flushleft]{threeparttable}
\usepackage{amssymb}
\usepackage{amsmath}
\usepackage{caption}
\usepackage{booktabs}
\usepackage{array}
\usepackage{paralist}
\usepackage{bigstrut}
\usepackage{multirow}
\usepackage{dcolumn}
\usepackage{siunitx}
\usepackage{graphicx}
\usepackage{tabularx}
\newcolumntype{Y}{{\centering\arraybackslash}X}

\begin{document}
\firstpage{1}

\subtitle{Phylogenetics}

\title[SPAGETI]{SPAGETI: Stabilizing Phylogenetic Assessment with Gene Evolutionary Tree Indices}
\author[Amiryousefi, A.]{Ali Amiryousefi\,$^{\text{\sfb 1,}*}$}
\address{$^{\text{\sf 1}}$Systems Oncology Unit, Faculty of Medicine, University of Helsinki, Helsinki, Finland \\}

\corresp{$^\ast$To whom correspondence should be addressed.}

\history{Received on X; revised on X; accepted on X}

\editor{Associate Editor: X}

\abstract{The standard approach to estimate species trees is to align a selected set of genes, concatenate the alignments and then estimate a consensus tree. However, individual genes contain differing levels of
evolutionary information, either supporting or conflicting with the consensus. Based on individual gene evolutionary tree, a recent study has demonstrated that this approach may result in incorrect solutions and developed the internode certainty (IC) heuristic for estimating the confidence of splits made
on the consensus tree. Although an improvement, this heuristic neglects
the differing rates of molecular evolution in individual genes. Here
I develop an improved version of this method such that each gene is proportionally weighted based on its overall signal and specifically with the imbalanced signal for each node represented with gene tree. \\
\textbf{Contact:} \href{ali.amiryousefi@helsinki.fi}{ali.amiryousefi@helsinki.fi}.\\
\textbf{Key words:} Gene tree, Species tree, Bipartition, Internode reliability, Tree reliability.\\
}

\maketitle
\section{Introduction}
Proposed on 1983, \citep{Felsenstein1983}, bootstrap and its theory for
assessing the reliability of a phylogenetic tree is still being argued
\citep{Holmes2003}. Despite nonrealistic assumptions of the method,
such as independence of the residues, the method is still commonly
being used as a measure of reliability of the phylogenetic tree.
Nowadays method has been applied also to consensus trees estimated
from concatenated alignments of genes. In this case, use of bootstrap
induces a bias due to
the fact that bootstrap samples from a large amount of sites always
produce very similar trees  \citep{Salichos2013}. As a result, bootstrap will falsely give
high confidence values for consensus trees from concatenated gene
alignments, even if the tree itself is incorrect.
One of the alternatives as a measure of nodes certainty of a consensus tree in the case of the availability of gene trees are based on the nodes bipartition idea first proposed by Salichos \citep{Salichos2013}. This
sounds appealing since it is making its judgment of the nodes
placements based on the topology of the gene trees. While still as a
generalization of this method \citep{Salichos2014} tries to gather even
more information from the gene trees, the the information related to
the molecular clock of each gene is neglected and the focus is only
placed on the bipartitions deduced from the gene trees. One immediate
improvement would be to integrated this information and tune up the
values of nodes certainty. There are at least two conceivable source
of information that could be exploited to make such
improvement. First, one can look at the genes molecular clock and
correspondingly weight the genes that bear more evolutionary
information, higher in their analysis. And second, it is crucial to
integrate the evolutionary distance of two clades caused by a specific
bipartition, as long as one is concerned with an index that relies on
the gene trees.
Here I propose a feasible method to take such information into account and provide a theory that is needed to support our method. 
\section{Bipartition indices}
Being concerned with a measure of reliability over a consensus tree, in specific on the node placements, one can focus on each gene tree. Efforts has been made to address this issue in different ways. For example \cite{Salichos2013} proposed a gene-support frequency(GSF) as well as internode certainty (IC) that relies on the bipartition observed on the topology of each gene tree for a specific determined node on the consensus tree. Given the maximum number of the internodes of a consensus tree $L$, each internode $o_k\in \mathcal{O}, k=1, \ldots, L$ will result on a bipartition $B=K|K^{'}$, where $K$ is the species set descendants to the node $o_k$ and $K^{'}$ is the rest of the species. 
\subsection*{Internode and tree certainty}
The ratio of the number of the genes that exhibit the same type of bipartition, $m_s$ to the total number of the genes, $m$ will account for the GSF. IC on the other hand, will look for the most frequent conflicting bipartition $m_{f_{1}}$, as well. Then form a proper probabilistic measure over $1=p_s + p_{f_{1}}$ with $p_s=\frac{m_s}{m_s+m_{f_{1}}}$ and $p_{f_{1}}=\frac{m_{f_{1}}}{m_s+m_{f_{1}}}$ and apply the modified version of the Shanon's notion of entropy as,
\begin{equation}
IC = 1 + p_s log_2(p_s) + p_{f_{1}}log_2(p_{f_{1}}). \label{IC}
\end{equation}
The conflicting bipartition is defined as the bipartition that has at least one species replaced between $K$ and $K^{'}$ and of course they are not confined to one form. \cite{Salichos2014} has extended this definition of IC to a more generalized form that is not only restricted to the most frequent conflicting bipartition but all, resulting in ICA, 
\begin{equation}
ICA = 1 + p_s log_{1+c}(p_s) + \sum_{i=1}^{c}p_{f_{i}}log_{c+1}(p_{f_{i}}), \label{ICA}
\end{equation}
where $c$ is the maximum number of conflicting bipartitions that has the overall frequency of $0.05$ or more. This cut-off value is imposed as a trade off between accuracy and computational expense.  We further note that in this case the proper probability measure is based on $1=p_s + p_{f_{1}} + \ldots, p_{f_{c}}$. Also note the parabolic shape of the above formulas where it takes it minimum zero when the $p_s=p_{f_{1}}$ and $p_s=p_{f_{c}}, \forall c,$ in IC and ICA, respectively. \cite{Salichos2014} propose negative signs for the rare cases when $p_s < p_{f_{.}}$.\\
Note that (\ref{IC}) will give a value for each internode on the consensus tree. Upon gathering all of these values for all the internodes and adding them together, the tree certainty (TC), will be obtained as, $TC=\sum_{k=1}^L IC_k.$ Furthermore, the definition of TCA based on (\ref{ICA}) has been proposed as the counterpart of TC, $TCA=\sum_{k=1}^L ICA_k$. These two stand as the comparative values between different consensus trees.
In the following section I describe the model that integrates the genetic information laid in the gene trees and their underlying distance matrices. This integration, will be in two layers based on the genes tree topology. In the first scenario the mean of each gene tree distance matrix would be a weighting scale of each gene on the major scale. The second case on the other hand, account for the mean distance of the occurred bipartition on the gene, and incorporation of these values into IC and ICA. 
\section{Gene signals}
Suppose that we have the set of species $S=\{s_{1}, s_{2}, \ldots, s_{n}\}$, and the $i$th, $i=1,\ldots, n,$ species having a set of  genes $G_{i}=\{g_{i1}, g_{i2}, \ldots, g_{im_{i}}\}$. For simplicity let's consider $m_{i}= m, \forall i,$ while  $j=1,\ldots, m$ is prevailing the genes for species. This will result in a complete matrix of genes $\mathcal{G}_{nm}=\{g_{ij}\}$. Furthermore, let $\mathcal{T}=\{\mathcal{\tau}_1, \ldots, \mathcal{\tau}_m\}$ and  $\mathcal{D}=\{\mathtt{D}_i, \ldots, \mathtt{D}_m\}$ denote the set of gene trees and distance matrices related to each of the genes, respectively. Furthermore let $\mathbf{M}=\{\mu_1, \ldots, \mu_m\}$ be the vector of scalar means of corresponding lower triangle matrices of $\mathcal{D}$. Having the consensus phylogenetic tree $\mathbf{T}$, the aim is to attach a value to each node as a measure of reliability of that node's correct placement on that tree.     
   
\subsection*{Genes weighting; Major treatment}
 
 In the definition of (\ref{ICA}), we can integrate the information of the molecular clock of a gene into account; those genes that have more evolutionary signals be weighted more in the analysis. One index reflecting such information is embed in $\mu_.$, such that the higher this value the more evolutionary information is expected to be contained in gene $g_{-.}$. So one natural update in the values of GSF and ICA will be the weighting each genes by this corresponding values and update the probability measure accordingly. 
With this regard, we can form the weighted gene support (WGS) as $\frac{\sum_{i\in s} \mu_i}{\sum \mu_i}$, where $s$ is the set of all the genes supporting the specific bipartition.  Furthermore, with defining $\sigma_s= \sum_{i \in s}\mu_i$ and  $\sigma_{f_{.}}=\sum_{i \in f_{.}}\mu_i$, the counterparts of the $p_s$ and $p_{f_{.}}$ for the IC can be updated as,
\begin{equation}
\pi_s=\frac{\sigma_s}{\sigma_s+ \sigma_{f_{1}}}, \quad \label{sigma}
\pi_{f_{1}}=\frac{\sigma_{f_{1}}}{\sigma_s+ \sigma_{f_{1}}}
\end{equation}
and for the ICA as,
\begin{equation}
\pi_s=\frac{\sigma_s}{\sigma_s+ \sum_{j=1}^{c} \sigma_{f_{j}}}, \quad \label{sigma2}
\pi_{f_{.}}=\frac{\sigma_s}{\sigma_s+ \sum_{j=1}^{c} \sigma_{f_{j}}}. \quad
\end{equation}
These two will result in \textit{internode reliability} (IR) and for all (IRA) indices as follow,
\begin{equation}
IR = 1 + \pi_s log_2(\pi_s) + \pi_{f_{1}}log_2(\pi_{f_{1}}), \quad \label{IR}
\end{equation}
\begin{equation}
IRA = 1 + \pi_s log_{1+c}(\pi_s) + \sum_{i=1}^{c}\pi_{f_{i}}log_{c+1}(\pi_{f_{i}}).
\end{equation}

\subsection*{Bipartitions weighting; Minor treatment}

Moreover, by looking closer at bipartitions, we can infer more evolutionary information. The previous model is distinguishing between supporting and conflicting bipartitions and even has a concern on different types of bipartitions observed but still neglecting the different forms of the trees within each of these classes of bipartitions. It is indeed the case that all the supporting bipartitions are not equally supportive nor all the conflicting ones equally contradicting. To amend this situation we focus on the sub-topology of each gene trees governed with a specific bipartition as two mutually distinct sub trees. This in turn will cause a bipartition on the underlying distance matrix of that tree with corresponding mean values of their lower triangle values. For the $j$the gene we refer to $\{\tau_j^{K}, \tau_j^{K'}\}$, $\{\mathtt{D}_j^{K}, \mathtt{D}_j^{K'}\}$, and $\{\mu_j^{K}, \mu_j^{K'}\}$, as the bipartitions incurred on corresponding tree, distance matrix, and mean values, respectively. With setting the distance between two means of a specific bipartition as $d_j=|\mu_j^{K} - \mu_j^{K'}|$ we can form the interaction ratio as $\nu_j=d_j\mu_j$. This value shows the mixed effect  of overall gene signal that encompass both the gene importance \textit{per se} and its underlying structure imposed by a specific bipartition. Replacing the $\mu$ with $\nu$ as $\sigma_s= \sum_{i \in s}\nu_i$ and  $\sigma_{f_{.}}=\sum_{i \in f_{.}}\nu_i$ will make this minor treatment effective in calculating (\ref{sigma}) and (\ref{sigma2}) and correspondingly the values of IR and IRA.
\subsection*{Tree reliability (TR) and adopted IR}
Evidently now with the availability of the IR and IRA values one can talk about \textit{tree reliability} for the case of only most prevalent conflicting bipartition or many as $TR=\sum_{k=1}^L IR_k$, and $TRA=\sum_{k=1}^L IRA_k$, respectively.
Furthermore, since the interaction ratio. $\nu$, could be well close to zero in scenarios that either the gene is bearing weak signal or the partitioned clades are not distinctive, one can use the exponential of this value to ensure the minimum role of an observed bipartition.

%
%

\bibliographystyle{natbib}
\bibliography{Document}









\end{document}